\documentclass[12pt, a4paper]{article}
\usepackage[utf8]{inputenc}
\usepackage[T2A]{fontenc}
\usepackage[english]{babel}

\usepackage{feynmf}
\usepackage{tabularx}
\usepackage{booktabs}
\usepackage{footnote}

\usepackage{indentfirst}
\usepackage{amsmath, amssymb, mathrsfs}

\usepackage{lineno}

\usepackage{graphicx}

\usepackage{float}

\usepackage{caption}
\DeclareCaptionLabelSeparator{Sep}{\bf{:} }
\usepackage{csquotes}

\usepackage{array}
\usepackage{multirow}
\usepackage{hhline}

\usepackage{csquotes}
\usepackage{textcase}
\usepackage{xcolor,soul}

\usepackage{hyperref}
\hypersetup{colorlinks,urlcolor=blue,linkcolor=blue,citecolor=blue}

\usepackage{geometry}
\usepackage{setspace}
\newcommand{\cgtp}{\ensuremath{C_{\widetilde{G}+}/\Lambda^4}}
\newcommand{\cgtm}{\ensuremath{C_{\widetilde{G}-}/\Lambda^4}}

\newcommand{\cbw}{\ensuremath{C_{BW}/\Lambda^4}}
\newcommand{\cbb}{\ensuremath{C_{BB}/\Lambda^4}}
\newcommand{\cww}{\ensuremath{C_{WW}/\Lambda^4}}

\newcommand{\obw}{\ensuremath{\mathcal{O}_{BW}}}
\newcommand{\obb}{\ensuremath{\mathcal{O}_{BB}}}
\newcommand{\oww}{\ensuremath{\mathcal{O}_{WW}}}
\newcommand{\ogtp}{\ensuremath{\oper_{\widetilde{G}+}}}
\newcommand{\ogtm}{\ensuremath{\oper_{\widetilde{G}-}}}
\newcommand{\lagr}{\ensuremath{\mathcal{L}}}
\newcommand{\oper}{\ensuremath{\mathcal{O}}}

\newcommand{\zllg}{\ensuremath{Z(\ell\ell)\gamma}}
\newcommand{\zvvg}{\ensuremath{Z(\nu\nu)\gamma}}
\newcommand{\ETg}{\ensuremath{E_\text{T}^\gamma}}

\makeatletter
\def\maketitle{
\begin{center}\textbf{\Large{\@title}}
\par\vspace{2mm}\textbf{\@author}
\begin{singlespace}
\par$^a$National Research Nuclear University MEPhI, Moscow, Russia
\par$^b$A. Alikhanyan National Science Laboratory (Yerevan Physics Institute), Yerevan, Armenia\par
*E-mail: AESemushin@mephi.ru\par
$^\dag$E-mail: EYSoldatov@mephi.ru
\end{singlespace}\end{center}
}
\makeatother

\title{$CP$-violating effects of the neutral triple gauge couplings in $Z(\ell\ell)\gamma$ production at the LHC}
\author{Artur E. Semushin$^{a,b,}$*, Evgeny Yu. Soldatov$^{a,\dag}$, Anastasia S. Kurova$^{a}$}
\date{}

\begin{document}
\maketitle
\begin{abstract}
    The main goal of modern experiments in the high-energy physics area is to find deviations from the Standard Model, the theoretical framework which describes data well but is expected to be extended to a more general theory. The anomalous coupling approach provides an opportunity to look for a wide range of new physics effects in different experimental signatures thanks to its model independence. In this work, the neutral triple gauge couplings are considered in $\zllg$ channel, and an effective field theory is used to parametrize these couplings in the Lagrangian. Neutral triple gauge couplings are triple interactions between $Z$ bosons and photons, and some of them violate $CP$ symmetry. This work presents a study of $CP$-sensitive variables in the aforementioned channel using special angular variables and matrix-element-based optimal observables. Based on these variables, expected limits on the coupling parameters are set for the conditions of Run~II and Run~III at LHC experiments, demonstrating the possibility of studying the $CP$-violation using a neutral triple gauge coupling approach and special $CP$-sensitive variables.
\end{abstract}

\newpage
\section{Introduction}
The Standard Model (SM) of elementary particle physics describes experimental results obtained during the last 50 years in high-energy physics well, and no deviations from the SM (so-called new physics) have been found at the Large Hadron Collider (LHC) during Run~I and Run~II~\cite{ParticleDataGroup:2024cfk}. This leads to the fact that direct experimental searches for new particles become less prospective in the absence of significant energy growth. On the other hand, there are indirect approaches, which look for new physics by studying interactions of currently known particles. Thus, the latter ones are more suitable for the LHC Run~III and Run~IV conditions, since they provide a way to look for new physics at the energy scale beyond the accelerator constraints.

Search for anomalous couplings is an indirect, model-independent way to probe physics beyond the SM (BSM). It is based on looking for experimental events containing SM particles with anomalous kinematics compared to the SM predictions. This work studies neutral triple gauge couplings (nTGCs), which are forbidden in the SM at the tree level. Effective field theory (EFT)~\cite{Weinberg:1978kz, Degrande:2012wf} is used to parametrize these couplings in the Lagrangian. Free parameters of the EFT, Wilson coefficients, are usually constrained experimentally because of the absence of new physics manifestations. Setting such limits can provide the definition of the most likely region where the new physics is located, as well as the restriction of some SM extensions.

Anomalous couplings can violate $CP$ symmetry, and therefore, a search for them can provide a model-independent probe of $CP$ violation. Many experimental analyses study $CP$ properties of the anomalous couplings~\cite{ATLAS:2020evk, ATLAS:2022tan, ATLAS:2022vym, ATLAS:2023mqy, CMS:2024bua}. In the case of nTGCs, the analyses usually focus their searches on the $CP$-conserving contributions and do not study $CP$ violation~\cite{CDF:2011rqn, D0:2011tjg, CMS:2016cbq, ATLAS:2019xhj, ATLAS:2018nci, CMS:2021icx}. $CP$ violation in the nTGC sector has been probed in the ATLAS study of $ZZ(4\ell)$ production~\cite{ATLAS:2023zrv}, where the limits on the anomalous coupling parameters have been set based on $CP$ violating contributions. Moreover, $CP$-violating effects of nTGCs have been taken into account in the L3 study of nTGCs based on the $Z\gamma$ production in $e^+e^-$ collisions~\cite{L3:2004hlr}.

In this work, $CP$-violating nTGCs are studied using two kinds of special $CP$-sensitive variables. Reconstruction of such variables requires the presence of well-identified fermions in the final state. Therefore, for example, $\zvvg$ production at the LHC, which has a high sensitivity to $CP$-conserving nTGC contributions, has no sensitivity to $CP$-violating ones. Thus, this work uses $\zllg$ production at the LHC for the sensitivity studies and comparison of the two variables.

\section{Phenomenology of neutral triple gauge couplings}
New physics, which includes new heavy particles at an experimentally inaccessible energy scale $\Lambda$, appears at low energies as anomalous couplings of already known particles. Such anomalous couplings can be parametrized in the Lagrangian by EFT in the following way:
\begin{equation}
    \lagr = \lagr_\text{SM} + \sum\limits_{d>4} \lagr^{(d)},
\end{equation}
where $\lagr_\text{SM}$ is dimension-four SM Lagrangian, and $\lagr^{(d)}$ is the dimension-$d$ anomalous term. Each such term consists of a set of dimension-$d$ operators $\oper_i^{(d)}$ with corresponding Wilson coefficients:
\begin{equation}
    \lagr^{(d)} = \sum\limits_{i} \frac{C_i^{(d)}}{\Lambda^{d-4}} \oper_{i}^{(d)}.
\end{equation}
In the SM, the values of all the Wilson coefficients $C_i^{(d)}$ are zero. Experimentally, no nonzero values have been observed, but limits on the coefficients have been set. Wilson coefficients can be matched to the parameters of new physics models~\cite{Remmen:2019cyz, Ellis:2024omd}, and therefore, experimental limits on the Wilson coefficients constrain the parameter space of such theories.

The SM, as well as operators of five, six, and seven dimensions does not contain nTGCs at the tree level. Thus, the description of such interactions mostly arises in dimension-eight EFT operators. This work studies $CP$-violating effects, therefore, only $CP$-odd operators are considered. There are five such operators defined as follows~\cite{Degrande:2013kka, Ellis:2023ucy}:
\begin{align}
    & \obw = i \Phi^\dag B_{\mu\nu} \hat{W}^{\mu\rho} \{ D_\rho , D^\nu \} \Phi + \text{h.c.}, \label{eq:obw} \\
    & \obb = i \Phi^\dag B_{\mu\nu} B^{\mu\rho} \{ D_\rho , D^\nu \} \Phi + \text{h.c.}, \label{eq:obb} \\
    & \oww = i \Phi^\dag \hat{W}_{\mu\nu} \hat{W}^{\mu\rho} \{ D_\rho , D^\nu \} \Phi + \text{h.c.}, \label{eq:oww} \\
    & \ogtp = g^{-1} B_{\mu\nu} W^{a\mu\rho} \left( D_\rho D_\lambda W^{a\nu\lambda} + D^\nu D^\lambda W^a_{\lambda\rho} \right), \label{eq:ogtp} \\
    & \ogtm = g^{-1} B_{\mu\nu} W^{a\mu\rho} \left( D_\rho D_\lambda W^{a\nu\lambda} - D^\nu D^\lambda W^a_{\lambda\rho} \right). \label{eq:ogtm}
\end{align}
These operators are constructed out of the SM fields and parameters: $\Phi$, $B_{\mu\nu}$ and $\hat{W}_{\mu\nu}$ are the Higgs doublet, $U(1)$ and $SU(2)$ gauge field tensors, respectively, $g$ is the $SU(2)$ coupling constant, and $D_\mu$ is the covariant derivative.

\begin{figure}[h!]
    \centering
    \includegraphics[width=0.7\linewidth]{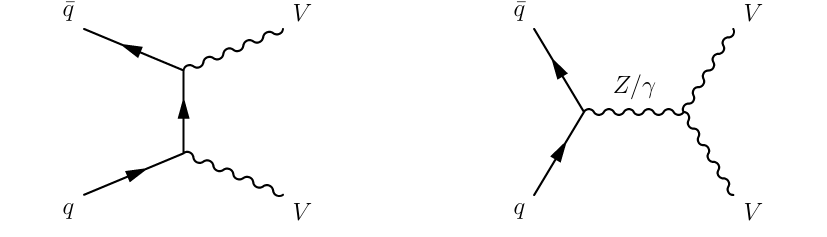}
    \caption{Feynman diagram examples for two neutral gauge boson productions at the LHC. The left diagram is allowed by the SM, whereas the right one contains a nTGC.}
    \label{fig:diagrams}
\end{figure}

The production of two neutral gauge bosons, such as $ZZ$ or $Z\gamma$, is usually a basis for studying nTGCs because of their sensitivity to such BSM manifestations~\cite{Biekotter:2021int}. Example diagrams of this process, containing zero or one nTGCs, are presented in Figure~\ref{fig:diagrams}.
Squared matrix element of the process, containing up to one nTGC, in case of one nonzero Wilson coefficient, can be written in the following form:
\begin{equation}
    |\mathcal{M}|^2 = |\mathcal{M}_\text{SM}|^2 + C/\Lambda^4 \cdot 2 \text{Re}\, \mathcal{M}_\text{SM}^\dag \mathcal{M}_\text{BSM}. \label{eq:decomposition}
\end{equation}
It consists of the SM term and interference (linear) term between SM-like and nTGC-like diagrams. The SM term is $CP$-conserving, whereas the interference term violates $CP$ in the case of $CP$-violating operators. It should be emphasized that the quadratic term is dropped in this work since it conserves $CP$. Additionally, the quadratic term at high sensitivity should be suppressed by the new physics energy scale ($\propto \Lambda^{-8}$) compared to the interference term. Furthermore, if the EFT expansion of $|\mathcal{M}|^2$ contains the quadratic term, then there might be missing terms. For instance, interference terms between dimension-ten (dimension-twelve) operators and SM have an order of $\Lambda^{-6}$ ($\Lambda^{-8}$) and can partially or totally cancel out the dimension-eight quadratic term. Thus, it is crucial to study linear-only expansion experimentally in addition to a common linear+quadratic expansion.

After phase-space integration, $CP$-violating contribution is zero and does not affect a total process cross section. However, differential distributions by special $CP$-sensitive variables can significantly change. Such variables allow one to separate positive and negative effects of $CP$ violation, and, therefore, are necessary for studies of $CP$-violating effects in nTGCs. This study uses $\zllg$ production as an example for variable construction and sensitivity study since it has a well-identified final state with fermions, which leads to a good $CP$ sensitivity.

\section{$CP$-sensitive variables}
\subsection{Angular variables}
\begin{figure}[h!]
    \centering
    \includegraphics[width=0.4\linewidth]{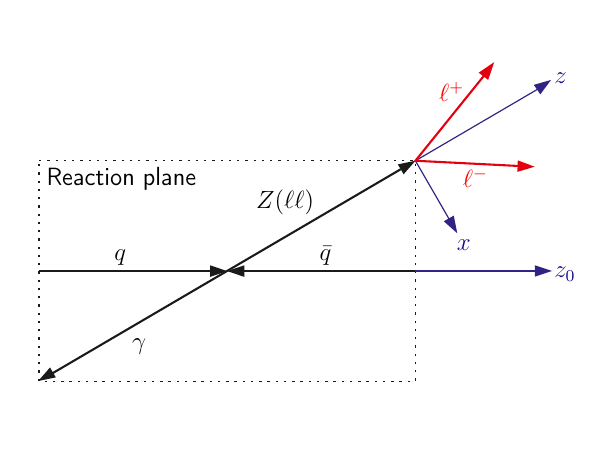}
    \hspace{0.05\textwidth}
    \includegraphics[width=0.35\linewidth]{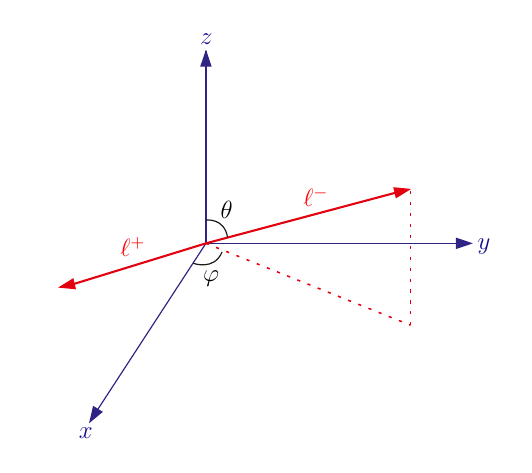}
    \caption{Scheme of the $\zllg$ production in the $\zllg$ rest frame with the new axes definitions (left) and $CP$-sensitive angle definitions in the $Z(\ell\ell)$ rest frame (right). In the left picture, all particles and axes lie on the reaction plane, excluding two leptons. The latter lie on the decay plane, which crosses the reaction plane along the $z$ axis. $z_0$ axis in this picture is the conventional axis directed along the beam pipe.}
    \label{fig:coordinates}
\end{figure}

Simple angular $CP$-sensitive variables can be constructed using the direction of motion of $Z$ boson decay products~\cite{Chang:1994cs}. For this purpose, a special reference frame should be used. In order to simulate the decay of $Z$ boson at rest, its rest frame is used. $Z$ boson decay can be described by two angles, $\theta$ and $\varphi$, and in this work they are polar and azimuthal angles of the direction of motion of the negatively charged lepton. These angles should be measured relative to the special coordinate system to be $CP$-sensitive, so the coordinate axes in the special reference frame are defined as follows: $z$ axis is set along the direction of motion of the $Z(\ell\ell)$ system in the $\zllg$ rest frame. $x$ axis is set so that it lies on the reaction plane, and its projection on the detector\footnote{Besides the special coordinate system for the study of $CP$ violation, this work also uses a conventional for LHC experiments coordinate system. Its origin is at the center of the detector; $x$, $y$, and $z$ axes are directed to the center of the LHC, upward and along the beam pipe, respectively.} $z$ axis has the same sign as new $z$ axis projection on the old $z$ axis. $y$ axis is totally defined by the $z$ and $x$ axes and set following the right-hand rule. Illustration of the special coordinate system is presented in Figure~\ref{fig:coordinates}.

\begin{figure}[h!]
    \centering
    \includegraphics[width=0.32\linewidth]{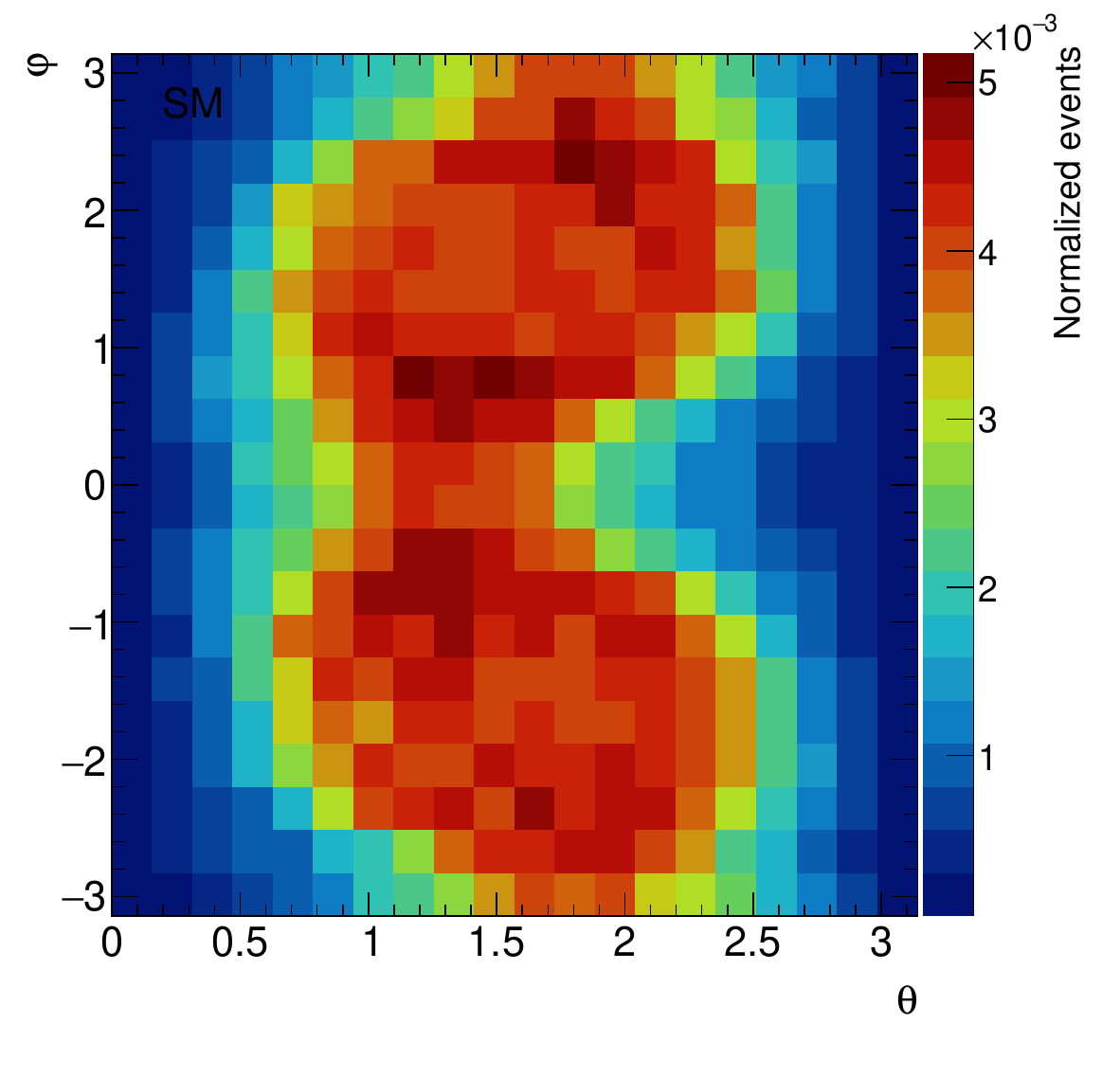}
    \includegraphics[width=0.32\linewidth]{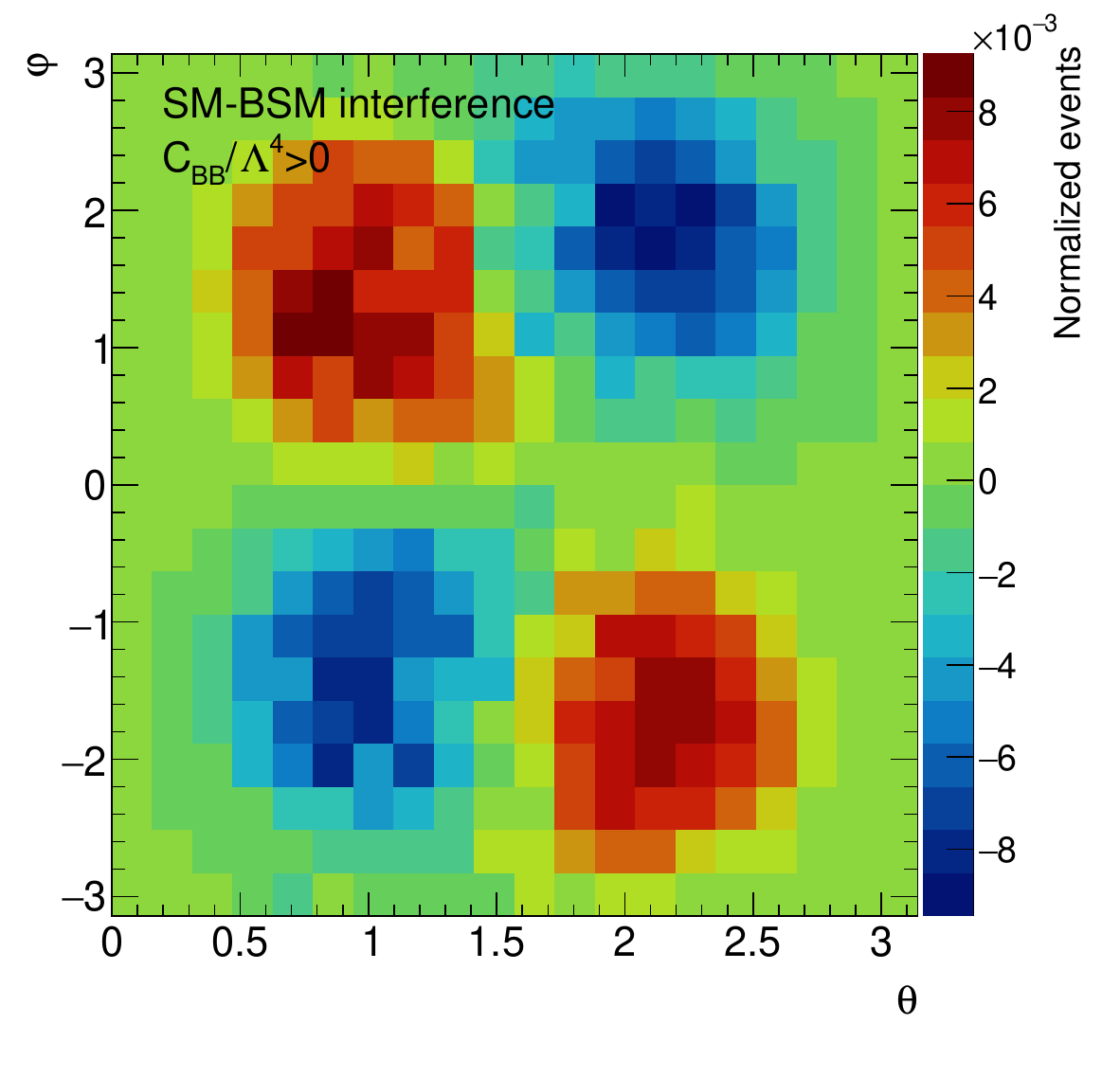}
    \includegraphics[width=0.32\linewidth]{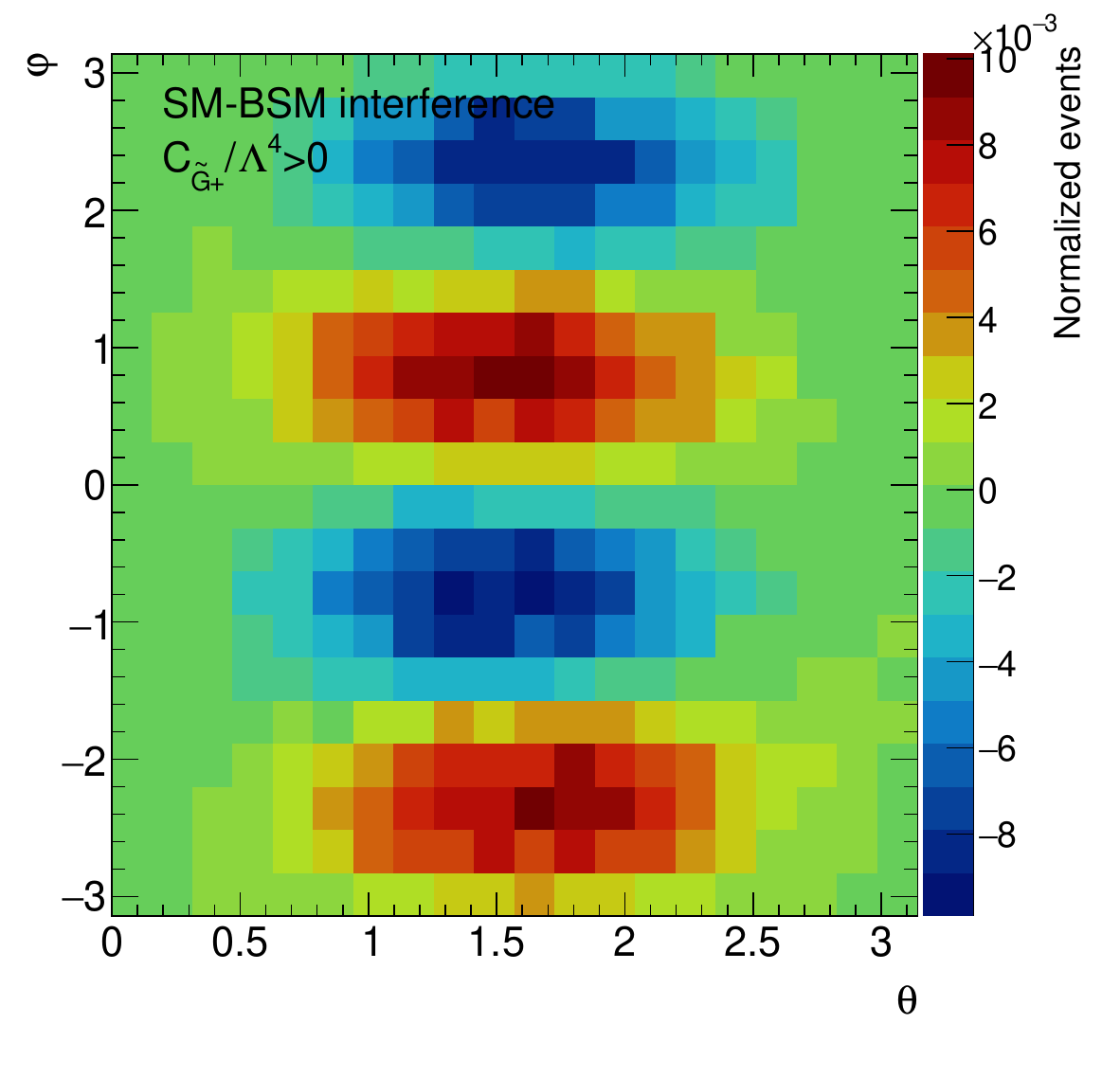}
    \caption{Normalized to unity two-dimensional distributions by $CP$-sensitive angles $\theta$ and $\varphi$: SM (left) and SM-BSM interference for $\cbb$ (center) and $\cgtp$ (right). Distributions for the other three coefficients are similar to the distribution for $\cbb$. For the interference terms, the normalization is done using absolute values of the number of events in bins.}
    \label{fig:ang_2D}
\end{figure}

\begin{figure}[h!]
    \centering
    \includegraphics[width=0.45\linewidth]{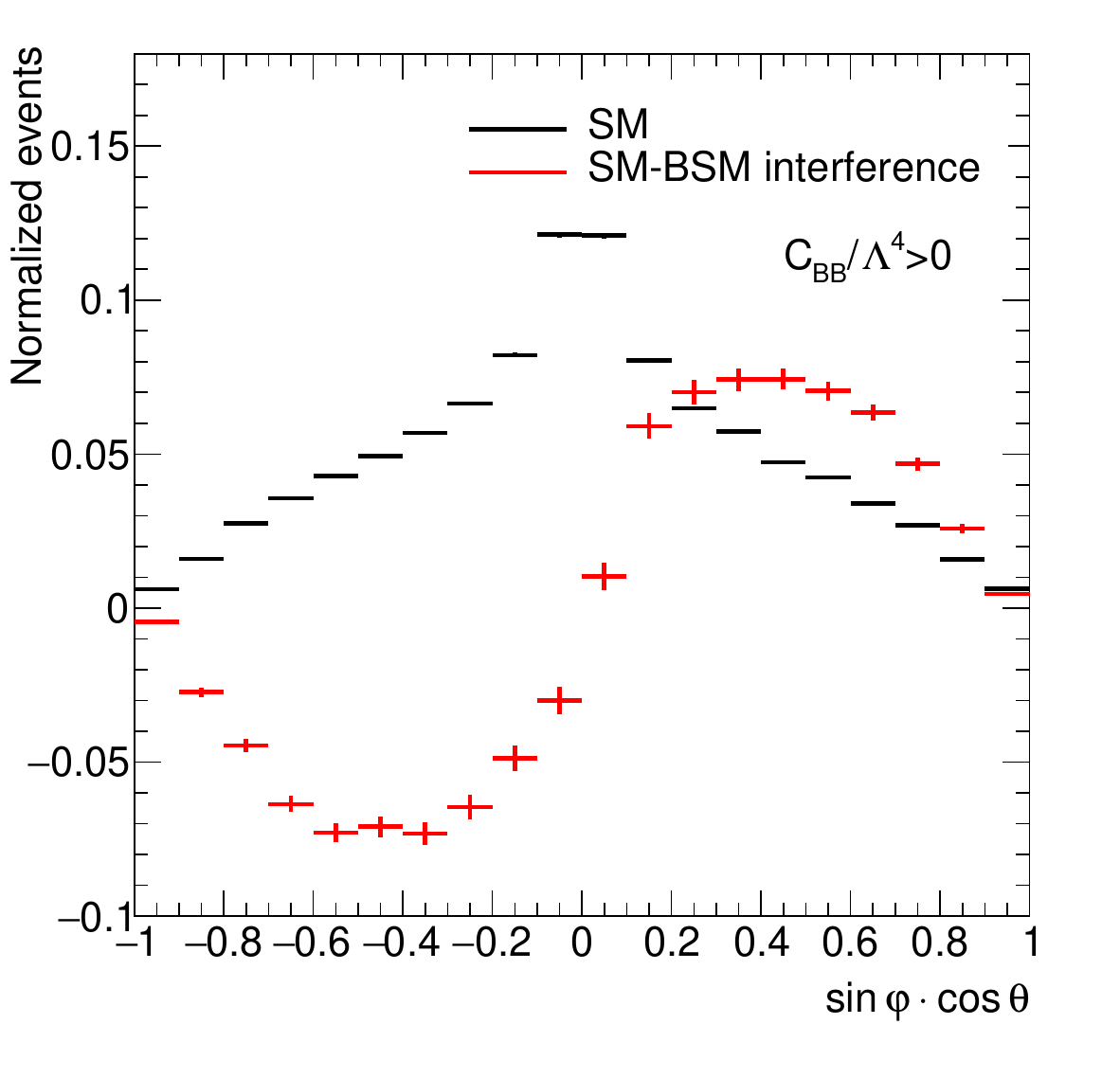}
    \includegraphics[width=0.45\linewidth]{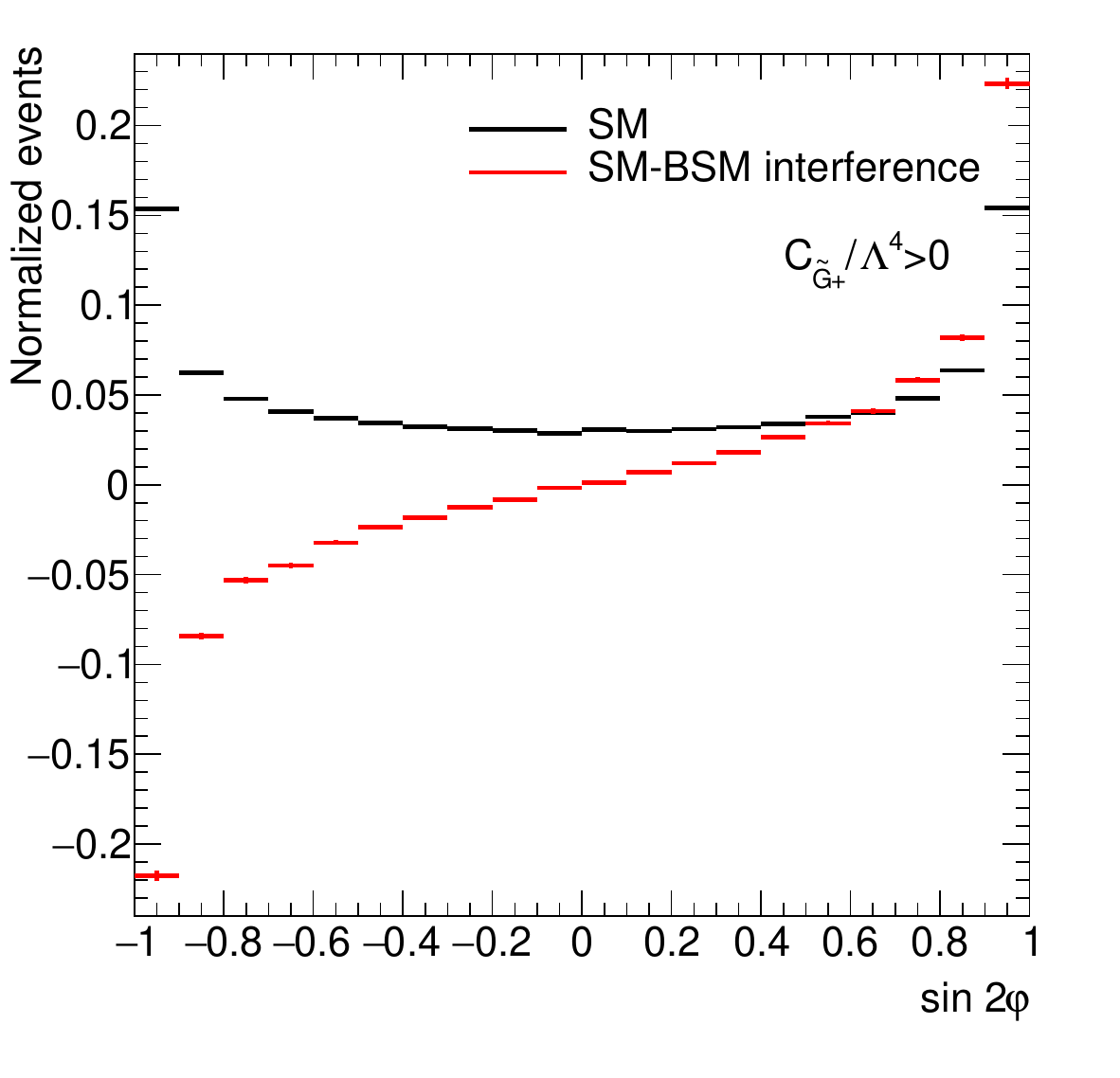}
    \caption{Normalized to unity distributions by $CP$-sensitive angular variables: $\sin \varphi \cos \theta$ for $\cbb$ (left) and $\sin 2\varphi$ for $\cgtp$ (right). For the interference terms, the normalization is done using absolute values of the number of events in bins.}
    \label{fig:decomp_ang}
\end{figure}

In general, $\theta$ and $\varphi$ are correlated, so they should be studied simultaneously to construct the $CP$-sensitive variable. Two-dimensional normalized distributions by these angles for SM and interference terms of two operators are presented in Figure~\ref{fig:ang_2D}. The SM contribution is positive in each bin, whereas SM-BSM interference has positive and negative parts. Localization of these parts at the two-dimensional angle distributions allows constructing the $CP$-sensitive variables, which separate positive and negative interference contributions. Thus, for coefficients $\cbb$, $\cbw$, $\cww$, and $\cgtm$ variable $\sin \varphi \cos \theta$ is used~\cite{ATLAS:2023zrv}. For coefficient $\cgtp$, the shape of a two-dimensional distribution is another, so a better separation power can be reached using another variable: $\sin 2\varphi$. The distributions of these variables can be found in Figure~\ref{fig:decomp_ang}.

\begin{figure}[h!]
    \centering
    \includegraphics[width=0.45\linewidth]{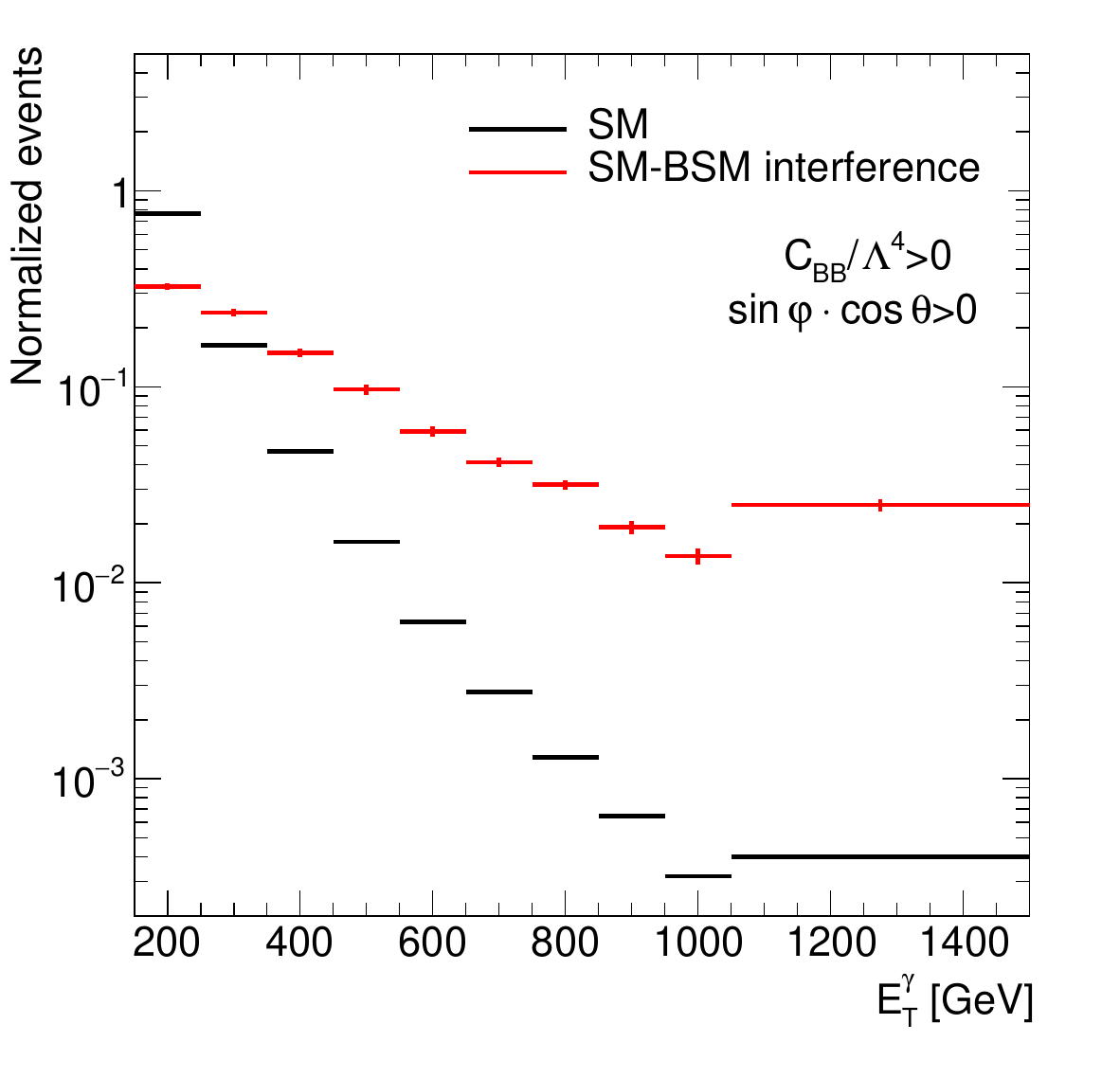}
    \includegraphics[width=0.45\linewidth]{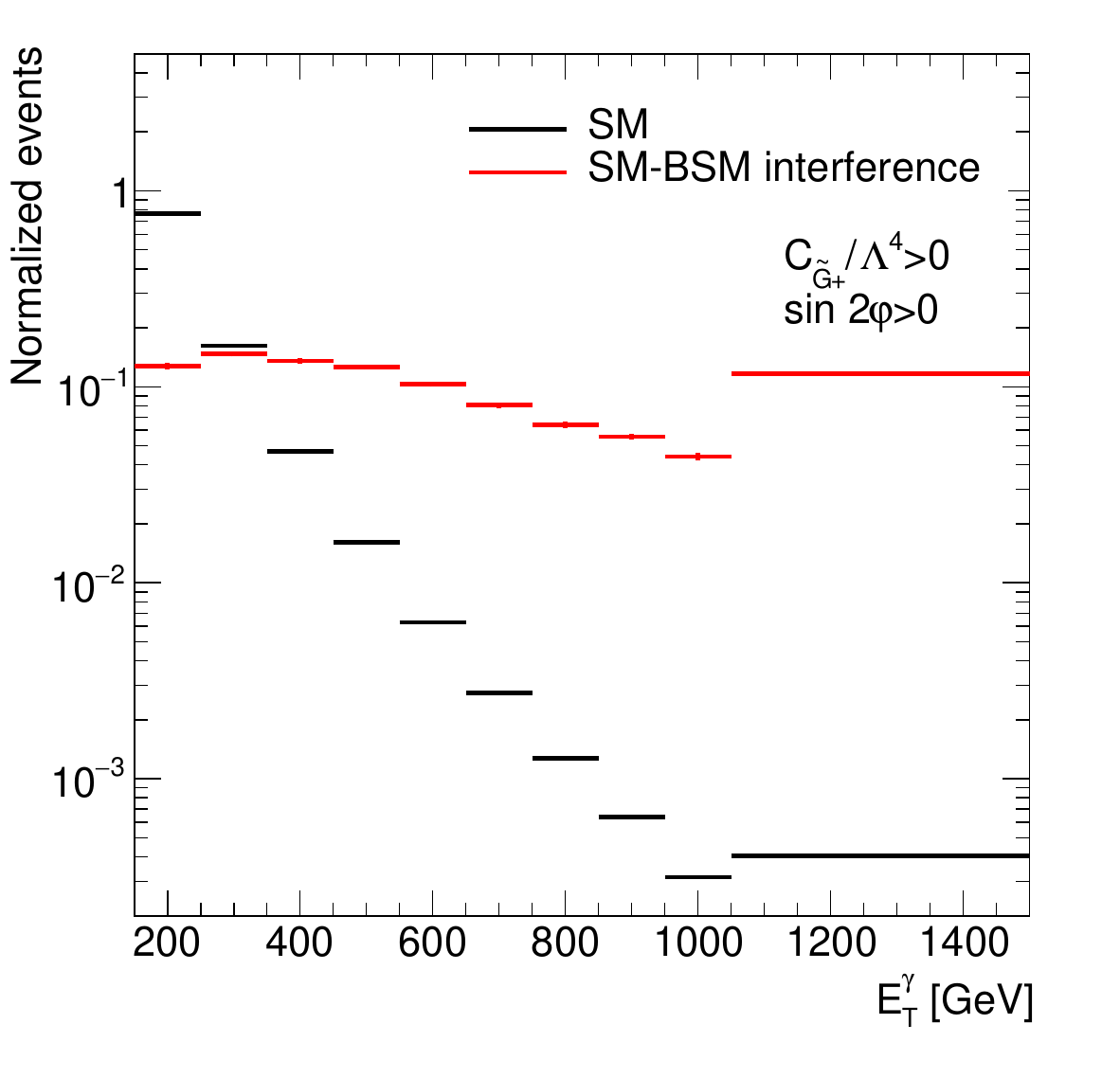}
    \caption{Normalized to unity distributions by photon transverse energy for SM and interference terms for $\cbb$ (left) and $\cgtp$ (right). Requirements $\sin \varphi \cos \theta >0$ (left) and $\sin 2\varphi >0$ (right) are applied to the distributions so that the interference term is positive. The last bin includes all overflow events.}
    \label{fig:decomp_ang_ptg}
\end{figure}

The aforementioned angular variables separate positive and negative interference contributions well, and their analyticity is their advantage. The regions of highest sensitivity of these variables are $|\sin \varphi \cos \theta| \gtrsim 0.2$ and $|\sin 2\varphi| \gtrsim 0.9$. Moreover, BSM contributions have another dependence on center-of-mass energy compared to the SM ones. Therefore, in order to reach better sensitivity to nTGCs, one should use a sensitive energetic variable in addition to the angular $CP$-sensitive variable via, e.g., cut or categorization. An example of nTGCs-sensitive energetic variable in $\zllg$ production is photon transverse momentum $\ETg$. Distributions by $\ETg$ are shown in Figure~\ref{fig:decomp_ang_ptg} for the positive part of the interference terms. At high $\ETg$ SM and interference contributions are well separated and the sensitivity is the highest.

\subsection{Optimal observables}
The optimal observable technique~\cite{Diehl:1996wm, Belyaev:2020hdo, Jahedi:2022duc} provides another way to construct the $CP$-sensitive variables. $CP$-sensitive optimal observable can be defined as the ratio of interference and SM parts of the process squared amplitude from Equation~\eqref{eq:decomposition}:
\begin{equation}
    OO = \frac{2 \text{Re} \, \mathcal{M}_\text{SM}^\dag \mathcal{M}_\text{BSM}}{|\mathcal{M}_\text{SM}|^2}.
\end{equation}
Such a definition can be understood as a likelihood ratio. Thus, such a variable is expected to have perfect sensitivity to nTGCs, combining sensitivity from $CP$-sensitive angular variables and energetic ones in the most optimal way. The optimal observable method was used in the nTGC study by the L3 Collaboration~\cite{L3:2004hlr}.

For the $pp$ collisions, components of optimal observable should be calculated using parton matrix elements $\mathcal{M}_\text{SM}^{ij}$ and $\mathcal{M}_\text{BSM}^{ij}$:
\begin{align}
    & 2 \text{Re} \, \mathcal{M}_\text{SM}^\dag \mathcal{M}_\text{BSM} = \sum\limits_{i,j} f_i(x_1, Q^2) f_j(x_2, Q^2) 2 \text{Re} \, \mathcal{M}_\text{SM}^{ij~\dag} \mathcal{M}_\text{BSM}^{ij}, \\
    & |\mathcal{M}_\text{SM}|^2 = \sum\limits_{i,j} f_i(x_1, Q^2) f_j(x_2, Q^2) |\mathcal{M}_\text{SM}^{ij}|^2,
\end{align}
where $i,j$ are initial-state partons, $f_i(x,Q^2)$ is the density function of parton $i$, which depends on collinear momentum fraction $x$ and factorization scale $Q$. Collinear momentum fractions of partons coming from positive and negative LHC $z$-axis directions can be reconstructed as 
\begin{align}
    & x_1 = \frac{m_{\ell\ell\gamma}}{\sqrt{s}} e^{y_{\ell\ell\gamma}}, \\
    & x_2 = \frac{m_{\ell\ell\gamma}}{\sqrt{s}} e^{-y_{\ell\ell\gamma}},
\end{align}
where $\sqrt{s}$ is $pp$ center-of-mass energy, $m_{\ell\ell\gamma}$ and $y_{\ell\ell\gamma}$ are invariant mass and rapidity of the $\ell\ell\gamma$ system. The choice of the factorization scale depends on its definition in the matrix-element-level Monte Carlo (MC) event generator, which is used to produce the events with nonzero Wilson coefficients. In this work, it is set to half of the sum of final particle transverse masses.

\begin{figure}[h!]
    \centering
    \includegraphics[width=0.45\linewidth]{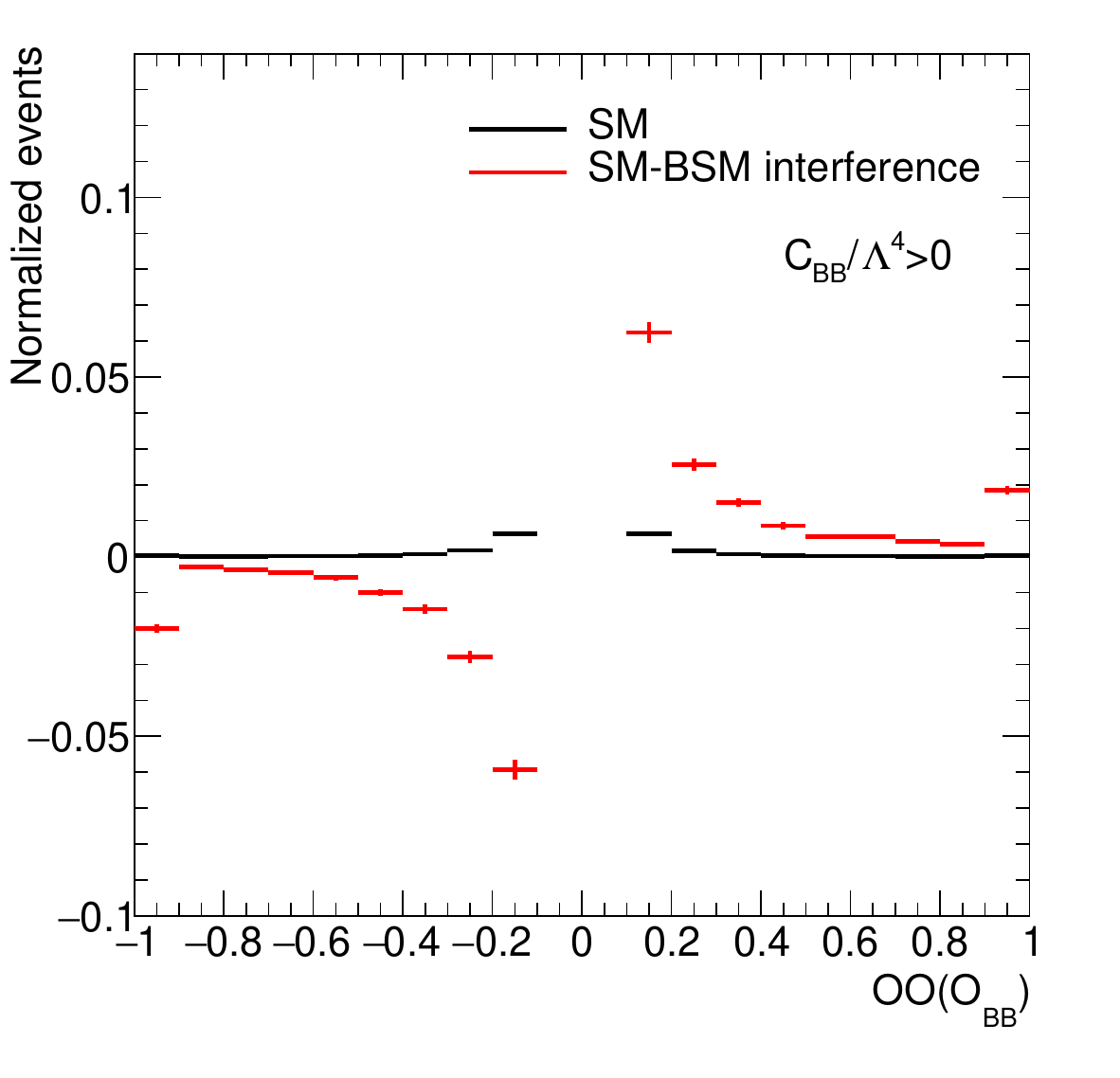}
    \includegraphics[width=0.45\linewidth]{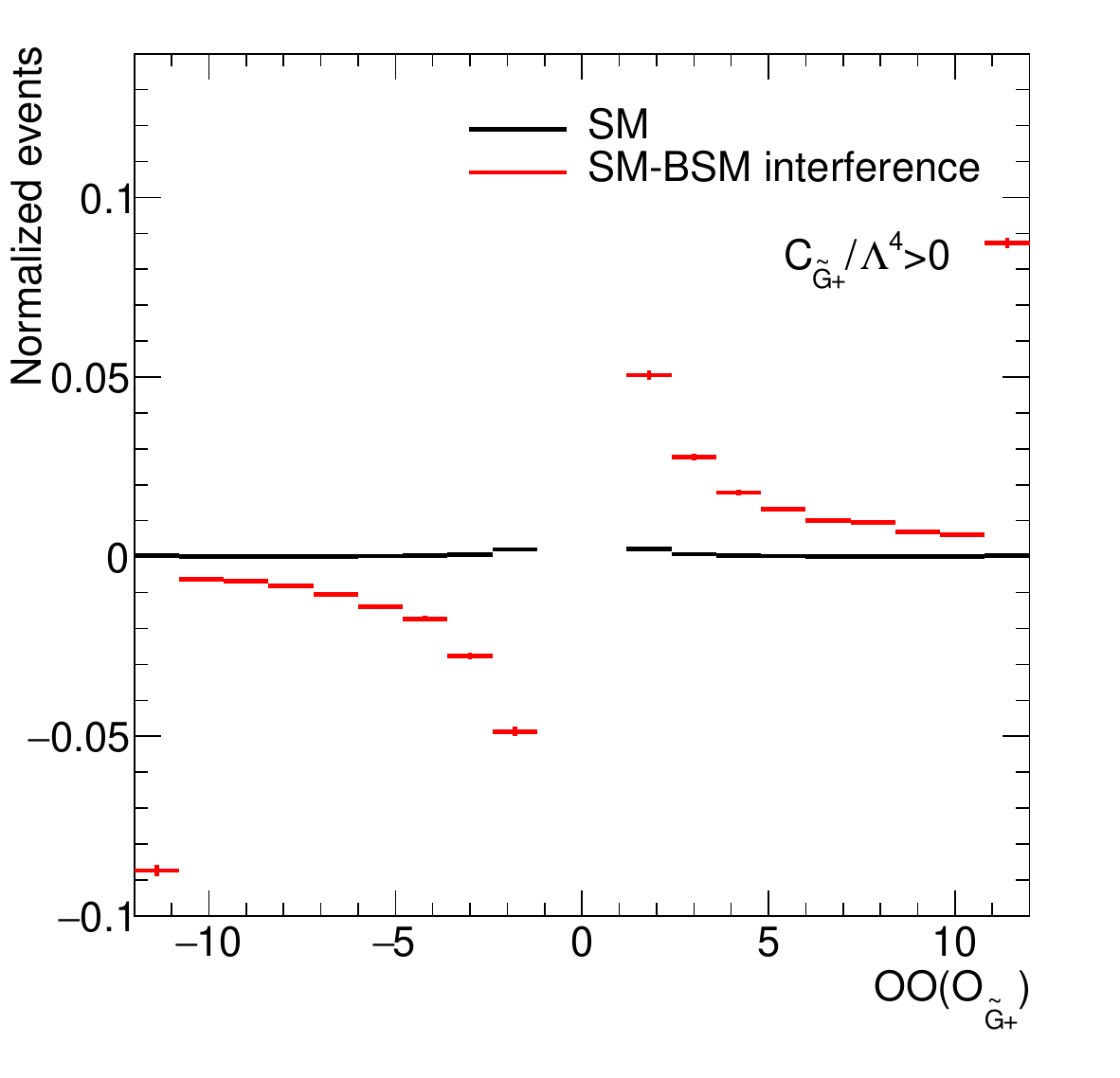}
    \caption{Normalized to unity distributions by $CP$-sensitive optimal observables for $\cbb$ (left) and $\cgtp$ (right). For the interference terms, the normalization is done using absolute values of the number of events in bins. Contents of two center bins are out of the plot range. The first and last bins include all underflow and overflow events correspondingly.}
    \label{fig:decomp_oo}
\end{figure}

\begin{figure}[h!]
    \centering
    \includegraphics[width=0.45\linewidth]{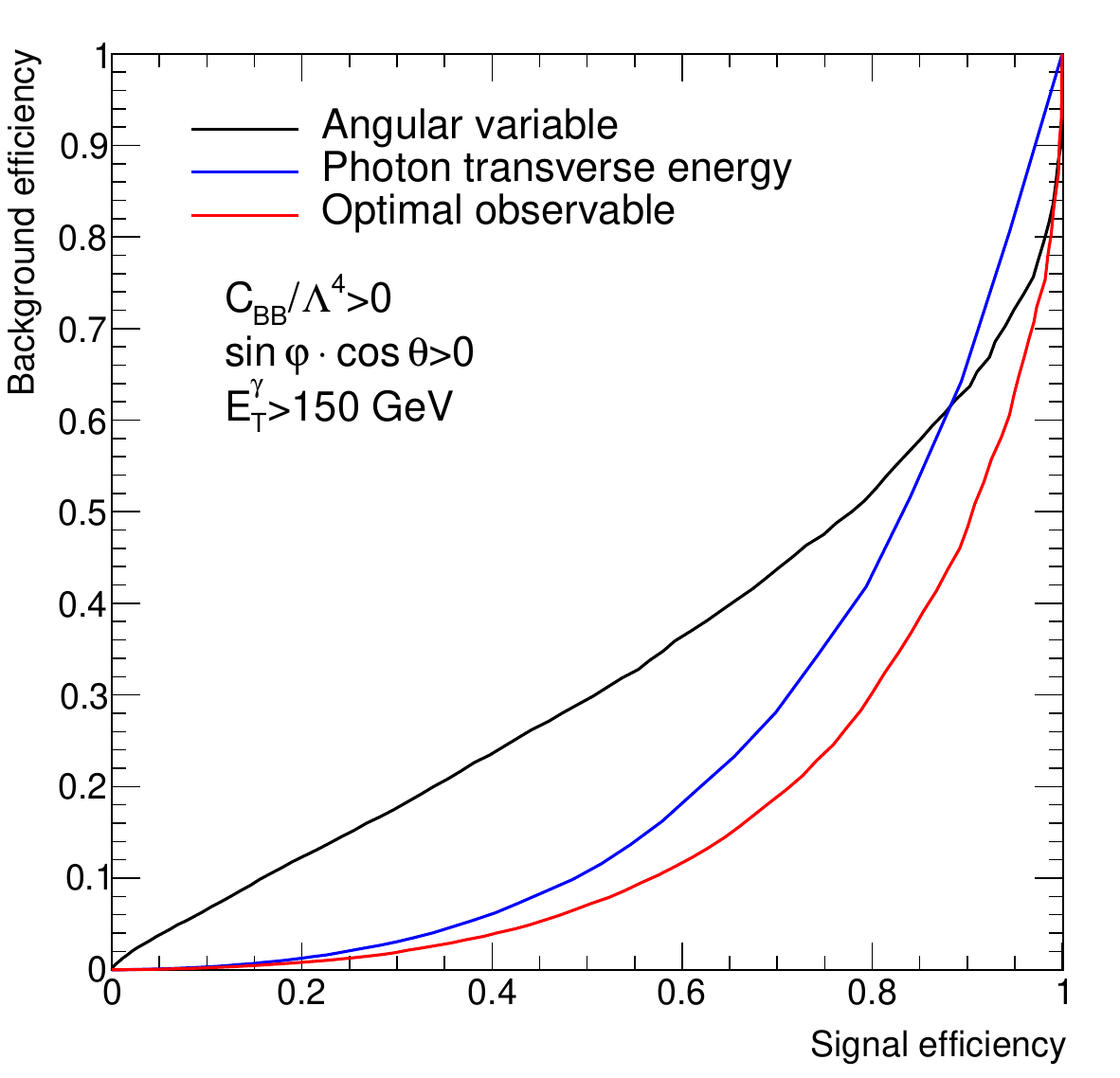}
    \includegraphics[width=0.45\linewidth]{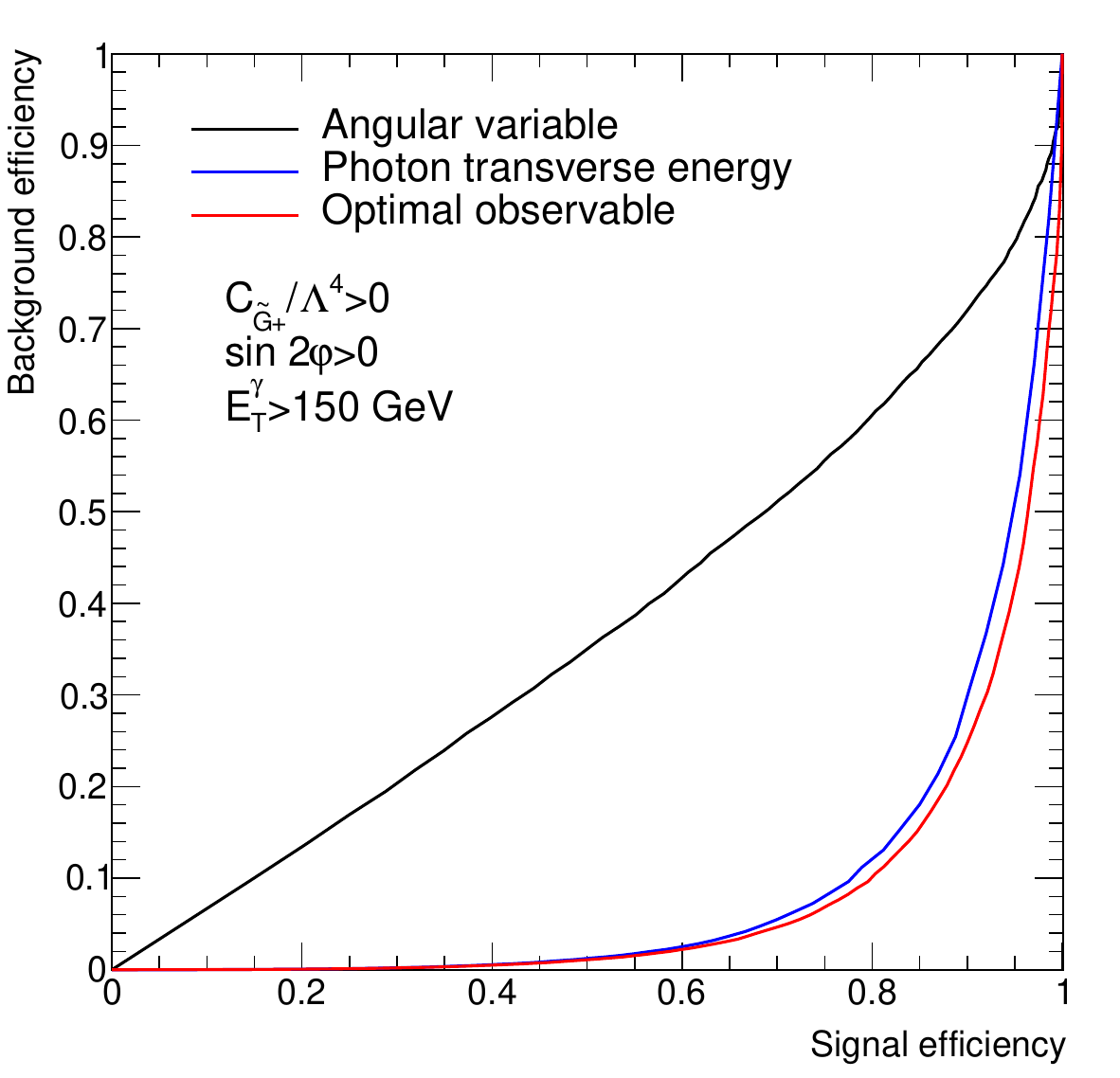}
    \caption{ROC curves for $\cbb$ at $\sin \varphi \cos \theta >0$, $\ETg>150$~GeV (left) and $\cgtp$ at $\sin 2\varphi >0$, $\ETg>150$~GeV (right). $\ETg$ threshold is explained further in the text.}
    \label{fig:roc}
\end{figure}

\begin{figure}[h!]
    \centering
    \includegraphics[width=0.45\linewidth]{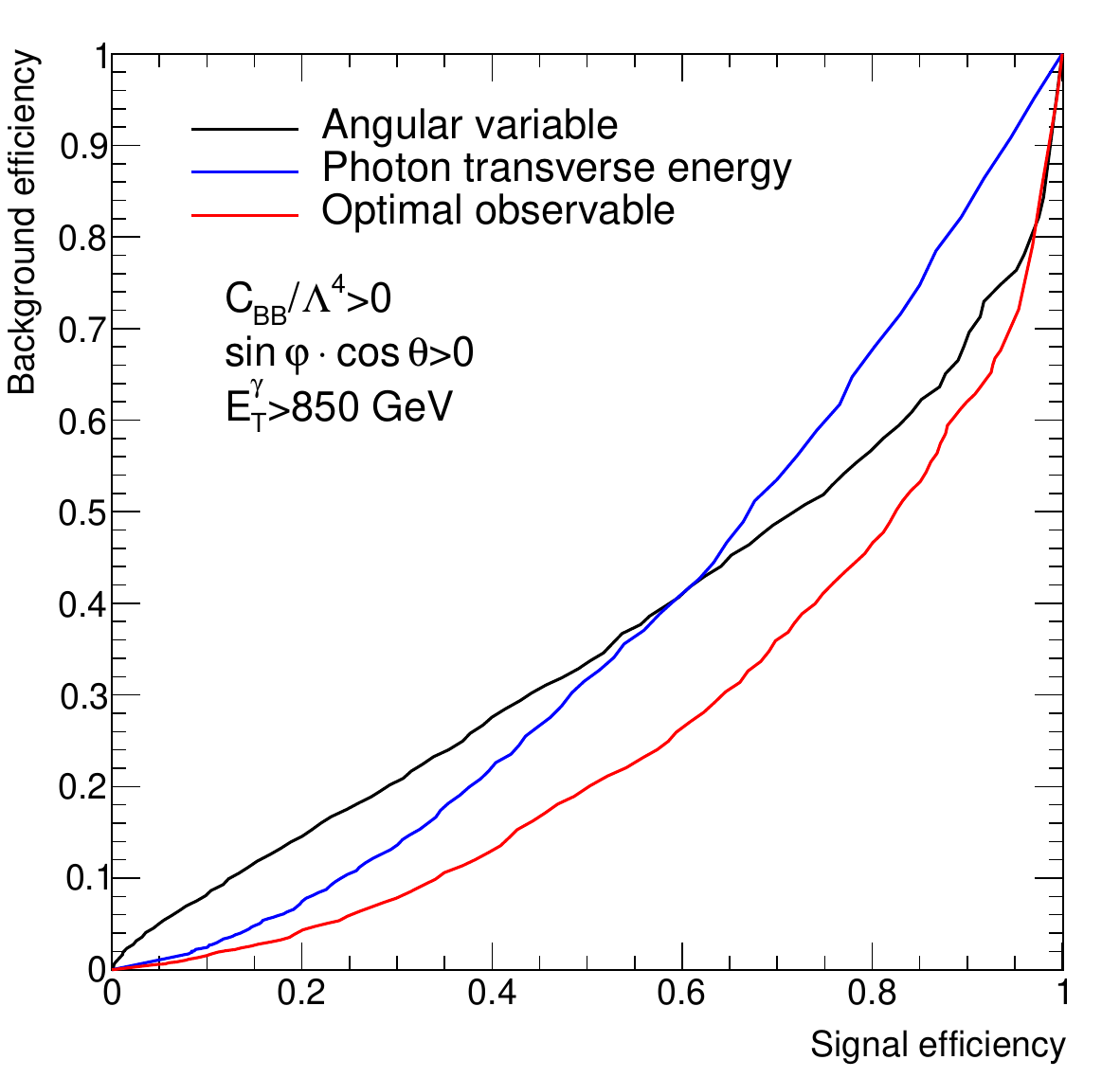}
    \includegraphics[width=0.45\linewidth]{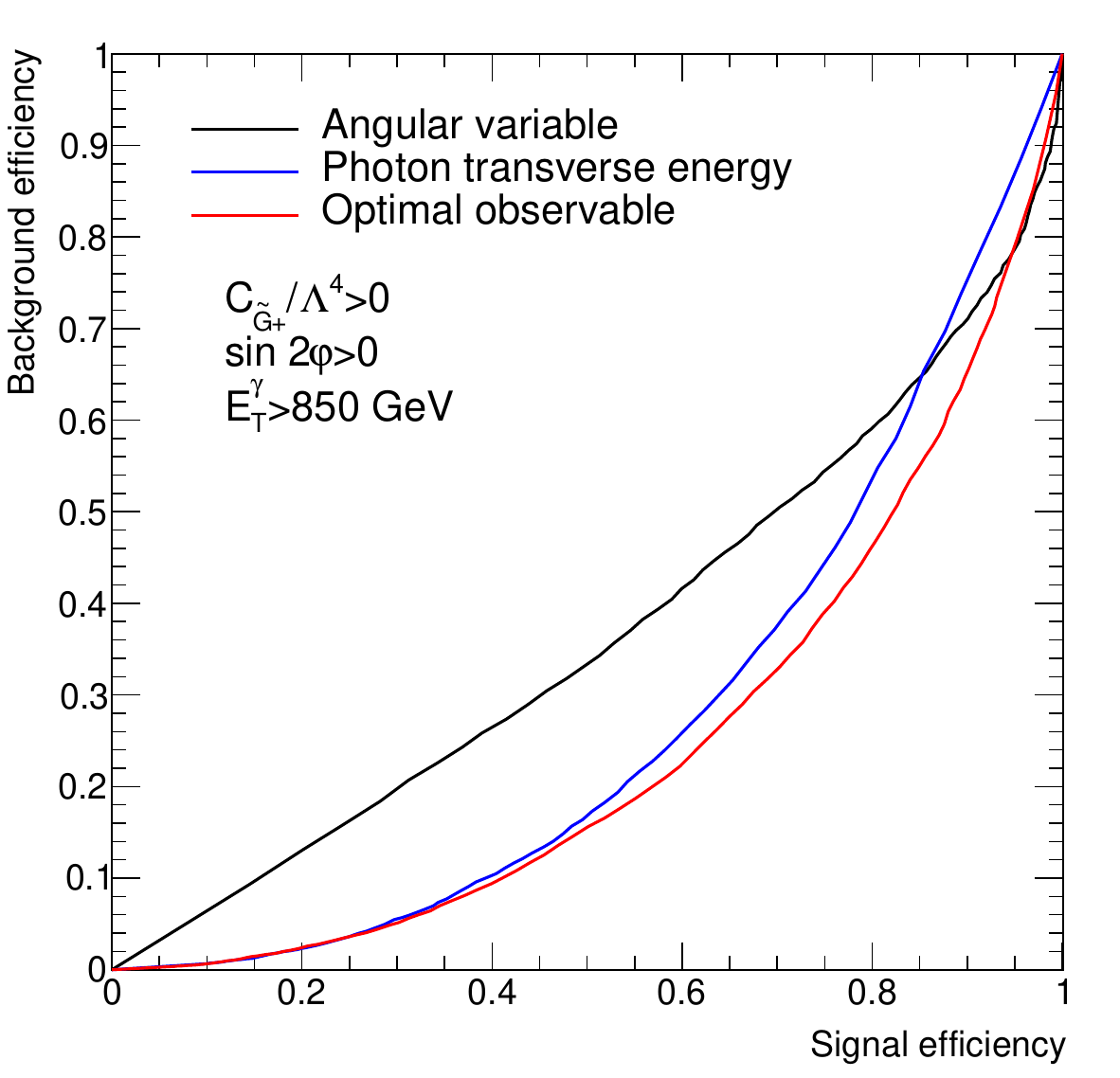}
    \caption{ROC curves for $\cbb$ at $\sin \varphi \cos \theta >0$, $\ETg>850$~GeV (left) and $\cgtp$ at $\sin 2\varphi >0$, $\ETg>850$~GeV (right). $\ETg$ threshold is explained further in the text.}
    \label{fig:roc_he}
\end{figure}

Optimal observable distributions for two operators are presented in Figure~\ref{fig:decomp_oo}. These variables allow one to reach the perfect sensitivity to nTGCs, especially at the left and right tails of the distributions. It should be emphasized that each operator has its own optimal observable. Receiver operating characteristic (ROC) curves for angular variables, photon transverse energy, and optimal observables are presented in Figures~\ref{fig:roc} and~\ref{fig:roc_he} for low and high photon transverse energies, respectively. They are created using the part of the distributions, where the interference term is positive. It can be seen that the performance of the optimal observable is the best. Angular variables show low sensitivity and are useful mainly for separation of positive and negative $CP$-violating contributions. Photon transverse energy has good sensitivity and is convenient to be used along with angular variables. Thus, nTGC limits based on the optimal observables are expected to be the most stringent.

\section{Sensitivity study for $Z(\ell\ell)\gamma$ production}
In order to study the sensitivity of $\zllg$ production to $CP$-violating effects of nTGCs, expected limits on five Wilson coefficients are calculated. The phase-space region used in this work is taken from the study of $\zllg$ production at LHC Run~II by the ATLAS Collaboration~\cite{ATLAS:2019gey}; the main event selection criteria are summarized in Table~\ref{tab:selections}. The additional criterion of $\ETg>150$~GeV is added similarly to ATLAS studies of $Z\gamma$ production~\cite{ATLAS:2018nci, ATLAS:2022nru} since low-energy contributions have negligible sensitivity to nTGCs. In this phase-space region, $\zllg$ production has some backgrounds which contamination is taken flat and equal to 18\% of the signal process SM event yield in accordance with the aforementioned study~\cite{ATLAS:2019gey}.
\begin{table}[h!]
    \centering
    \caption{Event selection criteria for the sensitivity study.}
    \label{tab:selections}
    \begin{tabular}{c}
        \hline
        1 opposite-charge same-flavor pair of leptons \\
        1 photon \\
        $p_\text{T}^{\ell_\text{leading}}>30$~GeV, $p_\text{T}^{\ell_\text{subleading}}>25$~GeV \\
        $\ETg>150$~GeV \\
        $m_{\ell\ell}>40$~GeV, $m_{\ell\ell\gamma}>182$~GeV \\
        \hline
    \end{tabular}
\end{table}

Events of $\zllg$ production in $pp$ collisions were generated at leading order using Monte Carlo event generator \texttt{MadGraph5\_aMC@NLO}~\cite{Alwall:2014hca}. SM and SM-BSM interference contributions were modeled separately to obtain the decomposed resulting samples. In order to improve the Monte Carlo statistics, events were generated separately by slices in photon transverse energy: $150 < \ETg < 300$~GeV, $300 < \ETg < 600$~GeV, and $\ETg > 600$~GeV. Universal FeynRules Output model~\cite{Degrande:2011ua}, including SM and $CP$-violating nTGC operators, was created in the \texttt{FeynRules} package~\cite{Alloul:2013bka} and used in the event generator. \texttt{Pythia8}~\cite{Sjostrand:2014zea} was used to model the parton shower, hadronization, and underlying event, and \texttt{Delphes3}~\cite{deFavereau:2013fsa} simulated the response of the ATLAS detector~\cite{ATLAS:2008xda} as an example of a typical detector at the LHC in this work. Matrix elements for optimal observables were calculated using stand-alone output in \texttt{MadGraph5\_aMC@NLO}. For both event generation and optimal observable computation parton density function set \texttt{NNPDF30\_nlo\_as\_0118} by NNPDF Collaboration~\cite{NNPDF:2014otw} was used via \texttt{LHAPDF} tool~\cite{Buckley:2014ana}. The SM contribution is known to have a significant next-to-leading order QCD correction due to the real emissions~\cite{Grazzini:2019jkl}. In order to take this into account, a $K$ factor was estimated as a ratio of the cross sections of the SM $\zllg$ production with up to two additional QCD jets and without QCD jets emitted. Its estimation of 2.5 was flatly applied to the SM prediction, and, after this, its agreement with the published ATLAS results~\cite{ATLAS:2019gey} was found to be reasonable.

\begin{figure}[h!]
    \centering
    \includegraphics[width=0.45\linewidth]{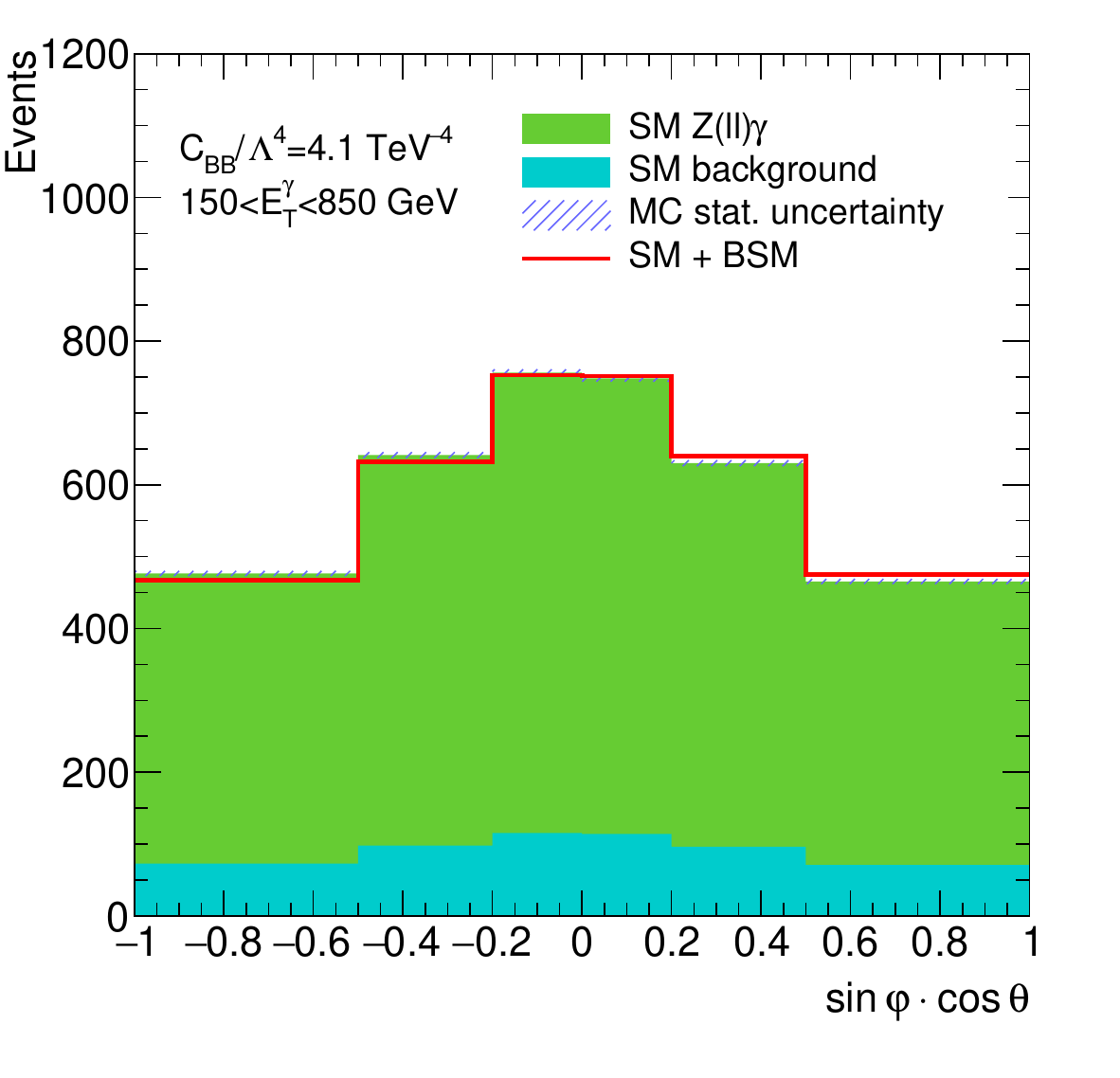}
    \includegraphics[width=0.45\linewidth]{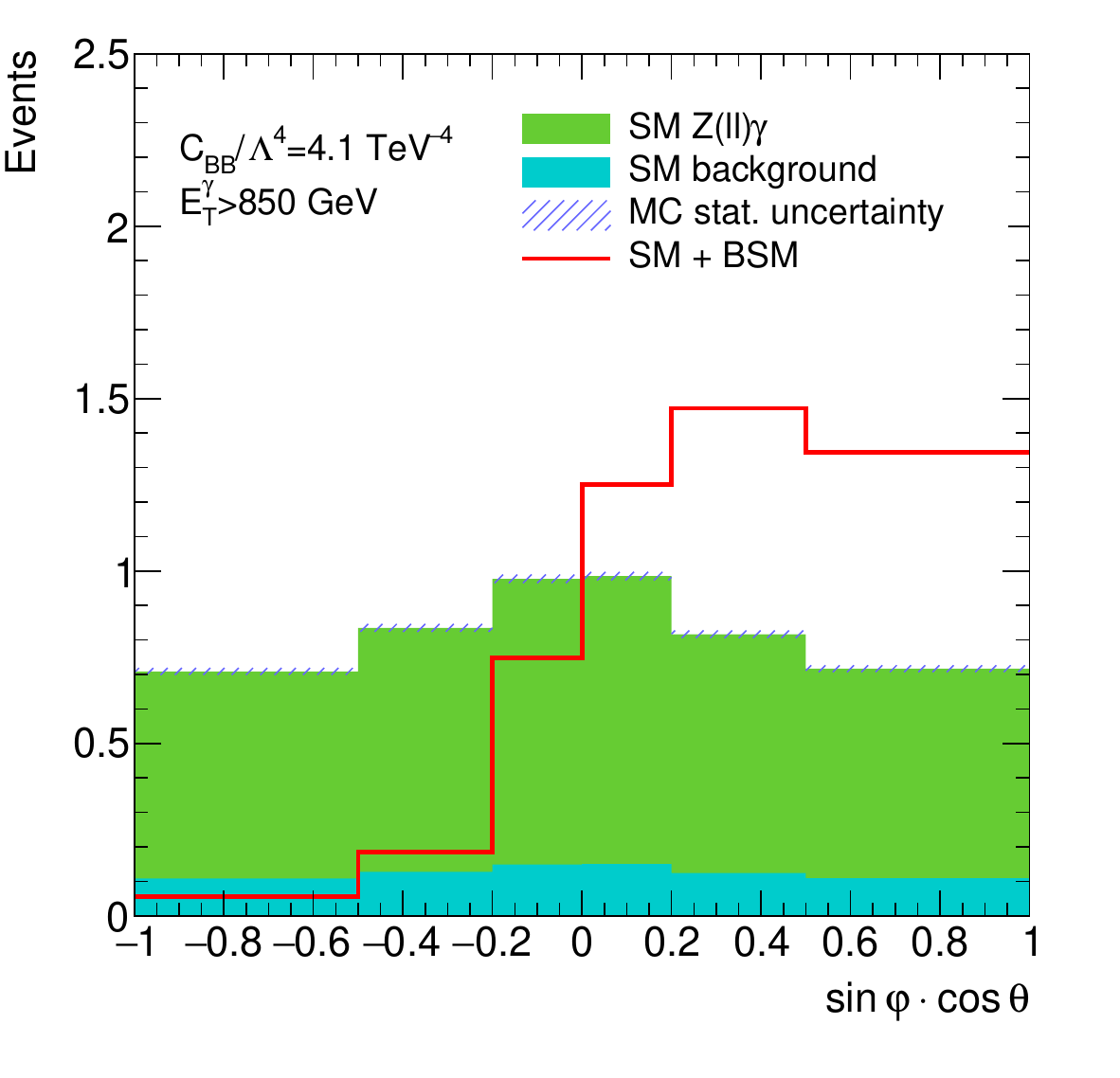}
    \caption{Distribution by $\sin \varphi \cos \theta$ for low-energy (left) and high-energy (right) slices. The case of Run~II integrated luminosity is presented. The dashed band shows the statistical uncertainty of MC modeling. The same distributions are used for $\cbw$, $\cww$, and $\cgtm$ coefficients.}
    \label{fig:model_ang_cbb}
\end{figure}

\begin{figure}[h!]
    \centering
    \includegraphics[width=0.45\linewidth]{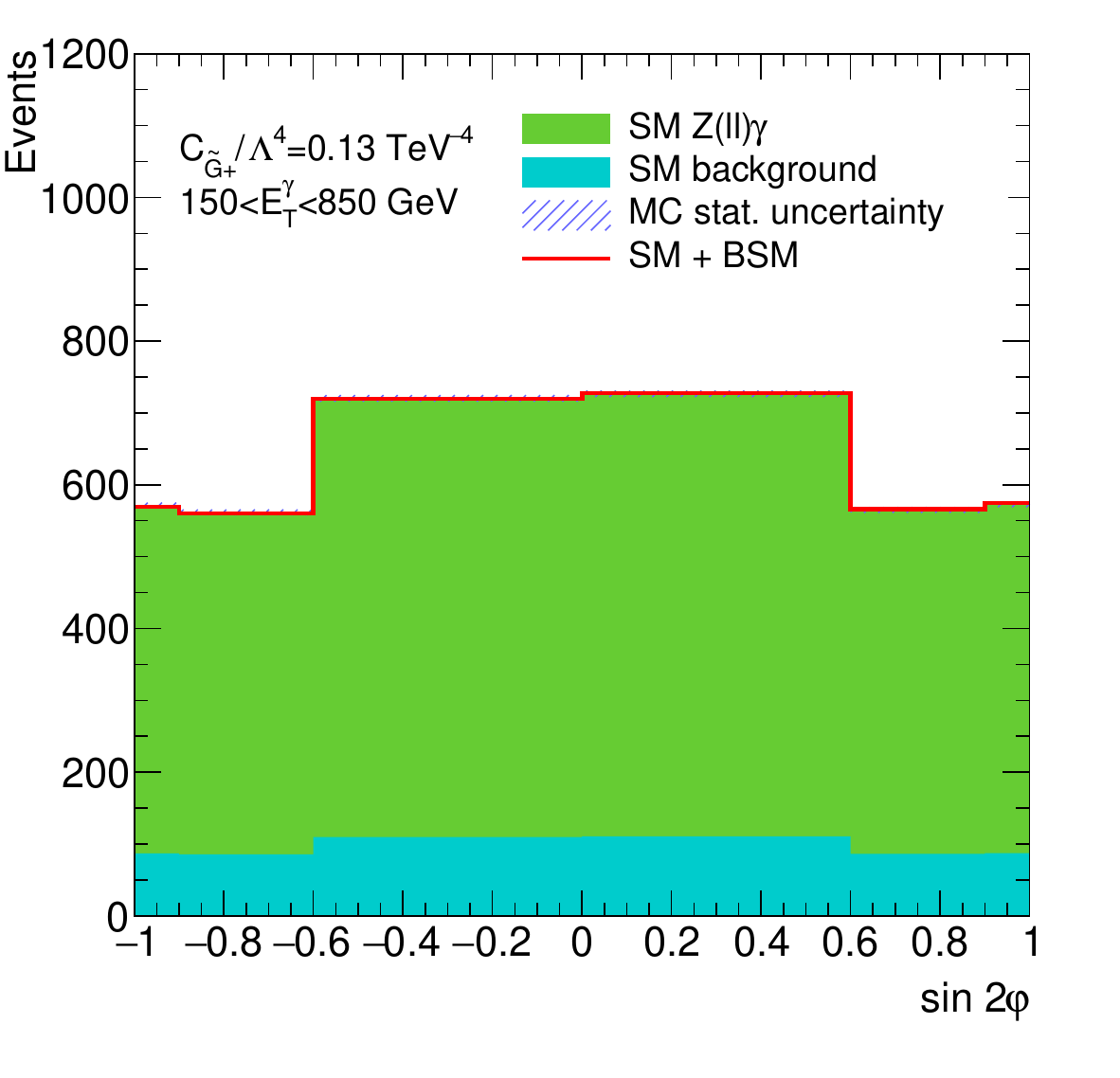}
    \includegraphics[width=0.45\linewidth]{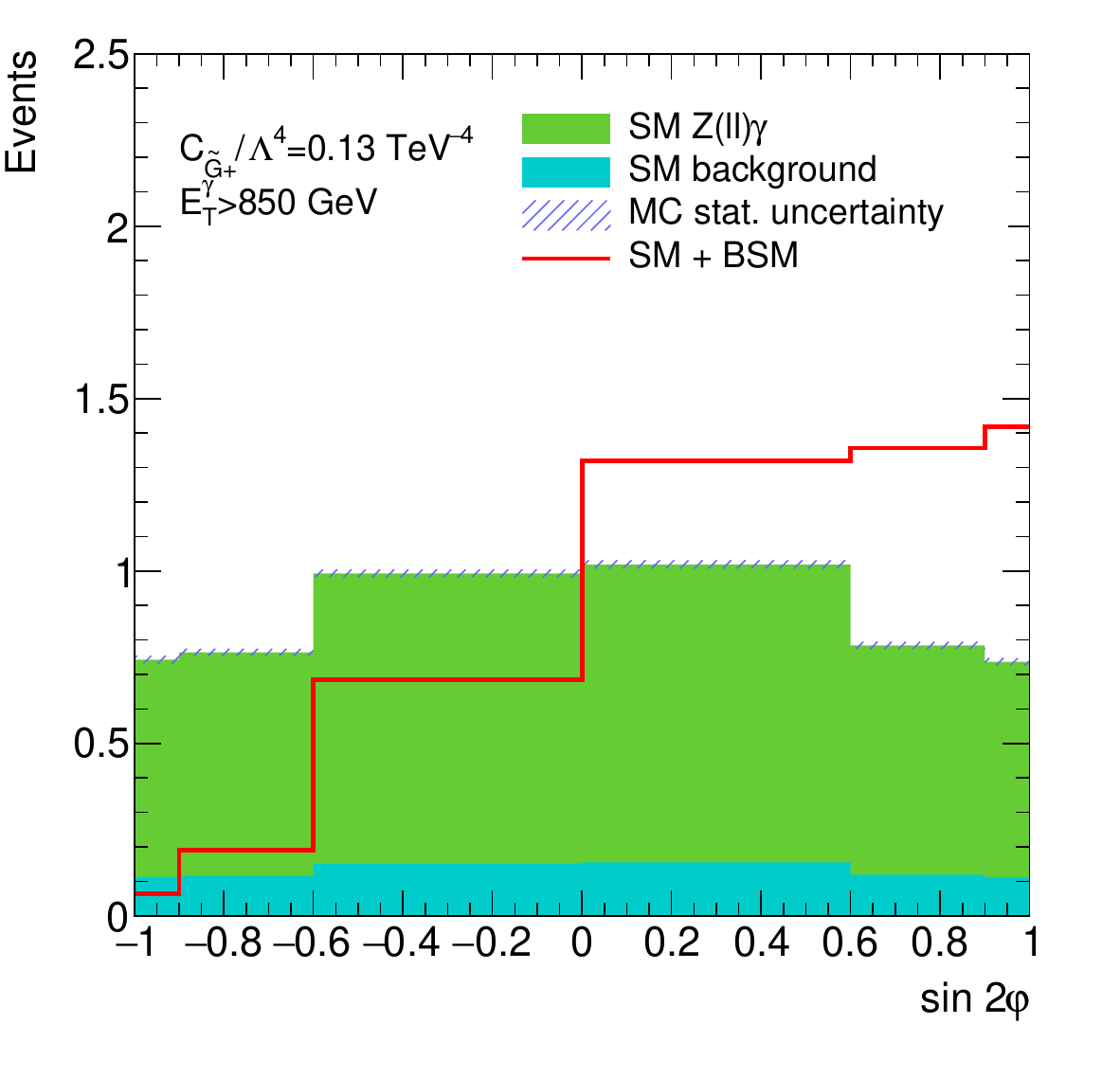}
    \caption{Distribution by $\sin 2\varphi$ for low-energy (left) and high-energy (right) slices. The case of Run~II integrated luminosity is presented. The dashed band shows the statistical uncertainty of MC modeling.}
    \label{fig:model_ang_cgtp}
\end{figure}

\begin{figure}[h!]
    \centering
    \includegraphics[width=0.45\linewidth]{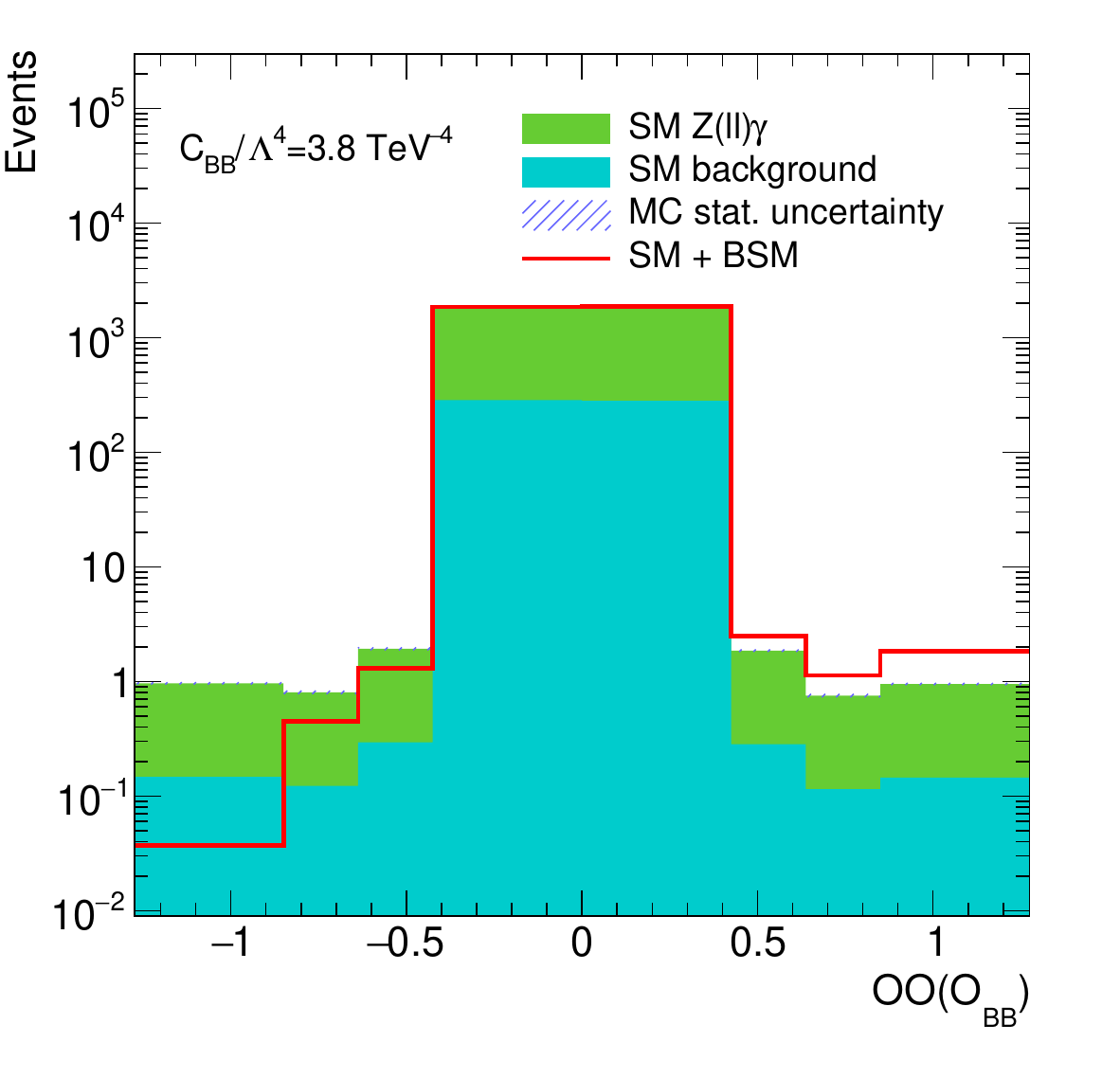}
    \includegraphics[width=0.45\linewidth]{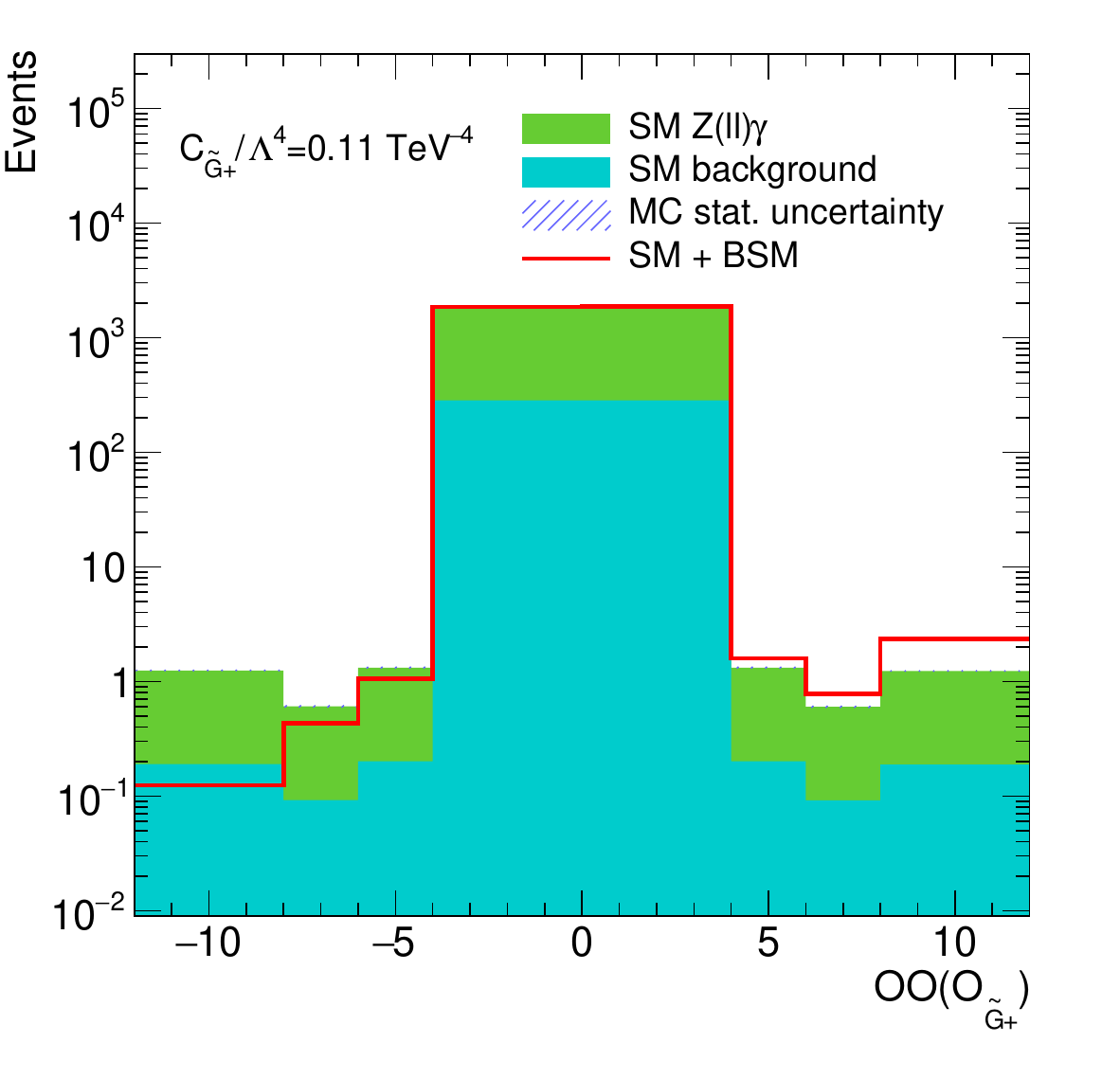}
    \caption{Distribution by optimal observable for $\cbb$ (left) and $\cgtp$ (right). The case of Run~II integrated luminosity is presented. The dashed band shows the statistical uncertainty of MC modeling.}
    \label{fig:model_oo}
\end{figure}

The limits on the Wilson coefficients are set using the frequentist approach with the likelihood-ratio-based test statistic, assuming its asymptotic distribution~\cite{Wilks:1938dza, Cowan:2010js}. The likelihood is constructed using distributions presented in Figures~\ref{fig:model_ang_cbb} and~\ref{fig:model_ang_cgtp} for angular variables and in Figure~\ref{fig:model_oo} for optimal observables. Distributions by angular variables are split into six bins, and two categories depending on $\ETg$ are used to improve the sensitivity. Distributions by optimal observables are split into eight bins. The optimization study is performed to ensure that more bins or categories do not lead to a significant increase of the sensitivity. The positions of the last bin for both variables, as well as the threshold separating categories, are set to reach the best experimental sensitivity. The following $\ETg$ thresholds between categories are set: 850~GeV for $\cgtp$, $\cbb$ and $\cbw$; 800~GeV for $\cgtm$; 900~GeV for $\cww$.

In addition to the MC statistical uncertainty of the process modeling, the additional systematic uncertainty of 10\% is added to the limit-setting procedure. Overall systematic uncertainties, which affect normalization only, do not have an impact on the resulting limits because of zero normalization of the $CP$-violating interference terms. Thus, additional systematic uncertainty is taken uncorrelated between bins, i.e., affecting both normalization and shape, and is expected to approximately estimate theoretical and experimental systematic uncertainties~\cite{ATLAS:2016qjc,ATLAS:2019gey}. The resulting limits on five Wilson coefficients set basing on the angular variable, and the optimal observable for LHC Run~II and projected Run~III integrated luminosities are presented in Table~\ref{tab:1Dresults}. The impact of the aforementioned systematic uncertainty on the limits is of 3--13\% depending on coefficient, variable, and luminosity. Optimal observables result in more stringent limits on $CP$-violating contributions of nTGCs. It should be emphasized that the optimization of binnings and threshold is done for the Run~II conditions, therefore, it is possible to improve the sensitivity for the Run~III conditions after the reoptimization.
\begin{table}[h!]
    \centering
    \caption{Expected limits [TeV$^{-4}$] on five Wilson coefficients set basing on $CP$-sensitive angular variables and optimal observables for Run~II and expected Run~III integrated luminosities.}
    \label{tab:1Dresults}
    \begin{tabular}{lllll}
        \hline
        \multirow{2}{*}{Coef.} & \multicolumn{2}{c}{140~fb$^{-1}$} & \multicolumn{2}{c}{300~fb$^{-1}$} \\
        & Ang. var. & Opt. obs. & Ang. var. & Opt. obs. \\ \hline
        $\cbb$ & [-4.2; 4.1] & [-3.9; 3.8] & [-3.3; 3.2] & [-3.0; 3.0] \\
        $\cbw$ & [-9.2; 8.1] & [-7.8; 7.4] & [-7.1; 6.4] & [-6.1; 5.8] \\
        $\cww$ & [-22; 21] & [-20; 19] & [-18; 17] & [-15; 15] \\
        $\cgtp$ & [-0.13; 0.13] & [-0.11; 0.11] & [-0.10; 0.10] & [-0.089; 0.090] \\
        $\cgtm$ & [-30; 33] & [-12; 12] & [-23; 25] & [-9.2; 9.0] \\ \hline
    \end{tabular}
\end{table}

For a majority of coefficients the optimal-observable-based limits are not much stricter than the angular-observable ones. It can be explained by the simplicity of the process considered, for which one can construct two common sensitive variables. In this case, usage of the common variables can be a more appropriate way, if one needs a good control under the phase space. However, more complicated processes might require a consideration of more sensitive variables at a time, whereas the optimal observable technique by construction perfectly combines all of them. It is important to note, that some limit values presented are very weak, so that the corresponding energy scales of new physics are below the $\ETg$ threshold we use. This contradiction and the fact that linear+quadratic-term-based experimental limits are usually much stricter can indicate that either the current experimental sensitivity is very low or the EFT has missing terms, which should significantly change the quadratic term contribution. This adds more significance to the studies of the linear-only EFT expansion.

\section{Conclusion}
This work studies $CP$-violating contributions of neutral triple gauge couplings at the LHC, using $\zllg$ production as an example. Often, the LHC experiments do not study $CP$-violating contributions and base their analyses on the $CP$-conserving contributions. $CP$ violation can be probed using interference between $CP$-violating BSM physics and the SM one. Pure BSM quadratic contributions can lead to significant improvement of the sensitivity, but do not allow studying $CP$ violation because of their $CP$-conserving nature and, thus, are not used in this work. Thus, this study shows an additional way to probe new physics in the nTGC sector and improve the limits.

$CP$-violating contributions were studied using two types of $CP$-sensitive variables: angular variables and an optimal observable. The angular observable $\sin \varphi \cos \theta$, measured for the negative lepton in a special reference frame, was shown to have good $CP$-sensitivity to Wilson coefficients $\cbb$, $\cbw$, $\cww$, and $\cgtm$. For coefficient $\cgtp$, the more appropriate variable is $\sin 2\varphi$. Usage of such angular variables requires application of additional energetic variables to increase the sensitivity. At the same time, it was shown that optimal observables provide a way to combine angular $CP$-sensitive and energetic variables into a powerful variable with the perfect sensitivity to nTGCs. Expected limits on five Wilson coefficients were set in this work, using only $CP$-violating contributions of nTGCs in $\zllg$ production. Limits based on the optimal observables are 8--63\% more stringent than the ones based on the angular variables with additional categorization in the energetic variable. Despite the fact that experimental limits, based on $CP$-conserving contributions, are basically more stringent than the ones based on $CP$-violating contributions~\cite{ATLAS:2023zrv}, the increase of the luminosity provides more sensitivity, and therefore, the ratio of the interference and quadratic term contributions rises. Therefore, it is possible to set more stringent limits by combining variables sensitive to $CP$-violating and $CP$-conserving contributions. Such a study can be applied to other channels, which has a sensitivity to $CP$ violation. Finally, it can be important at $ee$ colliders, such as the Circular Electron Positron Collider (CEPC)~\cite{CEPCStudyGroup:2023quu} and the International Linear Collider (ILC)~\cite{Behnke:2013xla}, which have lower center-of-mass energy, and where the interference term dominates over the $CP$-conserving quadratic one.

\section*{Acknowledgements}
The work was funded by the Ministry of Science and Higher Education of the Russian Federation, Project "New Phenomena in Particle Physics and the Early Universe" FSWU-2023-0073.

\bibliographystyle{JHEP}
\bibliography{references}

\end{document}